\documentclass[prl,reprint,superscriptaddress,preprintnumbers]{revtex4-1}
\pdfoutput=1
\usepackage{amsmath,amssymb,graphics}
\usepackage{verbatim}
\usepackage{graphicx}
\usepackage{color}
\usepackage{mathrsfs}
\usepackage{ragged2e}
\usepackage{float}
\usepackage{esint}
\usepackage[unicode=true,pdfusetitle,
bookmarks=true,bookmarksnumbered=false,bookmarksopen=false,
breaklinks=false,pdfborder={0 0 1},backref=false,colorlinks=true]
{hyperref}
\hypersetup{
	citecolor=black,linkcolor=black,urlcolor=black}

\def\equationautorefname~#1\null{eq.\,(#1)\null}
\usepackage[hang,flushmargin]{footmisc}
\allowdisplaybreaks
\makeatletter

\usepackage{etoolbox}
\apptocmd{\thebibliography}{\justifying\setlength{\leftskip}{7.4mm}}{}{}

\usepackage{relsize}

\makeatletter
\def\simgt{\mathrel{\lower2.5pt\vbox{\lineskip=0pt\baselineskip=0pt
			\hbox{$>$}\hbox{$\sim$}}}}
\def\simlt{\mathrel{\lower2.5pt\vbox{\lineskip=0pt\baselineskip=0pt
			\hbox{$<$}\hbox{$\sim$}}}}
\makeatother

\usepackage{changepage}

\newcommand{\be}{\begin{equation}}
	\newcommand{\ee}{\end{equation}}
\newcommand{\bea}{\begin{eqnarray}}
	\newcommand{\eea}{\end{eqnarray}}

\def\nn{\nonumber}

\def\ft#1#2{{\textstyle{\frac{\scriptstyle #1}{\scriptstyle #2} } }}
\newcommand{\w}[1]{\\[0.#1cm]}
\usepackage{color}
\definecolor{darkgreen}{rgb}{0,0.5,0}
\definecolor{darkred}{rgb}{0.5,0,0}
\definecolor{darkblue}{rgb}{0,0,0.6}
\definecolor{purple}{rgb}{0.4,.2,0.7}

\def\a{\alpha}
\def\b{\beta}

\def\d{\delta}

\def\g{\gamma}

\def\l{\lambda}
\def\m{\mu}
\def\n{\nu}

\def\r{\rho}
\def\s{\sigma}

\def\O{\Omega}

\begin{document}
	
	\title{The Surprising Effectiveness of Weyl Gravity in Probing Quantum Corrections to AdS Black Holes}
	
	\author{Liang Ma}
	\email{liangma@tju.edu.cn}
	\affiliation{Center for Joint Quantum Studies and Department of Physics,\\
		School of Science, Tianjin University, Tianjin 300350, China \\}
	
	\author{Peng-Ju Hu}
	\email{pengjuhu@tju.edu.cn}
	\affiliation{Center for Joint Quantum Studies and Department of Physics,\\
		School of Science, Tianjin University, Tianjin 300350, China \\}
	
	\author{Yi Pang}
	\email{pangyi1@tju.edu.cn}
	\affiliation{Center for Joint Quantum Studies and Department of Physics,\\
		School of Science, Tianjin University, Tianjin 300350, China \\}
	
	\author{Hong L\"u}
	\email{mrhonglu@gmail.com}
	\affiliation{Center for Joint Quantum Studies and Department of Physics,\\
		School of Science, Tianjin University, Tianjin 300350, China \\}
	\affiliation{{ Joint School of National University of Singapore and Tianjin University,\\
			International Campus of Tianjin University, Binhai New City, Fuzhou 350207, China}}

	\date{\today}
	
	%	\preprint{}

	\begin{abstract}
		Computing leading higher curvature contributions to thermodynamic quantities of AdS black hole is drastically simplified once the higher curvature terms are expressed in terms of powers of Weyl tensor by applying proper field redefinitions, avoiding  the usual complications caused by higher derivative Gibbons-Hawking-York (GHY) term or surface counterterms. We establish the method by computing the Euclidean action of general rotating AdS black holes in five dimensional quadratic curvature theories with or without supersymmetry and verifying the results numerically.  Our result is the state of the art for charged rotating AdS black holes in five dimensional minimal gauged supergravity including corrections from all three supersymmetric curvature squared terms.  Our approach facilitates precision tests in the AdS/CFT correspondence and should be applicable in diverse dimensions.

	\end{abstract}
	
	%	\pacs{??? ... ???}
	
	\maketitle
	\allowdisplaybreaks

	%%%%%%%%%%%%%%%%%%
	%\section{Introduction}
	\textit{Introduction.}---The Weyl squared action has played a versatile role in our pursuit of a quantum theory of general relativity. In four dimensions, it admits a convergent Euclidean functional integral \cite{Tomboulis:1983sw} and enjoys renormalizability and asymptotic freedom \cite{Fradkin:1981iu}. Interestingly, the quantum fluctuations break the local scale invariance inducing the Einstein-Hilbert term with a calculable Newton's constant \cite{Adler:1982ri,Zee:1980sj}. In the context of (A)dS/CFT correspondence, Weyl gravity modified by a purely topological contribution from a Gauss-Bonnet term turns out to be equivalent to renormalized Einstein gravity at tree level \cite{Maldacena:2011mk}, with suitable boundary conditions chosen \cite{Lu:2011ks, Hell:2023rbf}. The equivalence between Weyl gravity and Einstein gravity underlies the construction of the widely studied critical gravity \cite{Lu:2011zk} and admits generalizations also in higher dimensions \cite{Lu:2011ks, Anastasiou:2020mik}. The principal reason that Weyl gravity has not received general acceptance is because the fourth-order derivatives lead to ghostlike excitations in the linearized theory. However, the violation of tree level unitarity  may be cured by invoking the Lee-Wick mechanism \cite{Lee:1970iw,Cutkosky:1969fq,Hasslacher:1980hd}  or adopting the $PT$ symmetric inner product \cite{Bender:2007wu,Mannheim:2021oat}. In the full theory, there is the zero energy theorem \cite{Boulware:1983td} stating that the exact asymptotically flat solutions in Weyl gravity all have zero energy rendering the ghosts confined in the nonlinear theory at large distances.
	
	The aim of this article is to unveil another instrumental role of Weyl gravity in the framework of effective field theory of quantum gravity. Since the work of Gibbons and Hawking \cite{Gibbons:1976ue}, numerous endeavours have been devoted to compute the Euclidean action of spacetime which plays an important role in the study of black hole thermodynamics, holography and quantum cosmology. For general higher derivative gravities, the task becomes much more difficult since one encounters equations of motion of higher order in partial derivatives.  Built upon previous work \cite{Reall:2019sah}, we find that computation of the leading higher curvature contributions to the Euclidean action of AdS black hole is drastically simplified once the higher curvature terms are expressed in terms of powers of Weyl tensor by applying proper field redefinitions \footnote{By accident, it was noticed in \cite{Gubser:1998nz} that a specific combination of two terms quartic in Weyl tensor enjoyed similar properties.}. The computation of on-shell Euclidean action boils down to simply evaluating the uncorrected solution in the higher derivative action, but without having to concern about the complicated higher derivative Gibbons-Hawking-York (GHY) or surface counterterms which would be required in the ordinary approach. In particular, the new approach is convenient for studying Kerr-AdS  black holes since to obtain the higher derivative corrections to the Euclidean action, one only needs to evaluate a bulk integral which has no preference on the choice of coordinates that may affect the induced metric on the conformal boundary \cite{Gibbons:2004ai}. Our new approach has been verified for static charged AdS black hole in a related work \cite{Hu:2023gru} announced recently. (See \cite{Xiao:2023two} for the background subtraction approach.) We have also confirmed the finiteness of Euclidean action for general asymptotically locally AdS spacetime (AlAdS) in Einstein gravity perturbed by a Weyl squared term. In this letter, employing our method, we obtain, for the first time, corrections to thermodynamics of Kerr-AdS  black holes from generic quadratic curvature terms. This result is also verified numerically.  Equipped with the powerful new method, we also revisit the leading higher derivative corrections to charged rotating black holes in five dimensional minimal gauged supergravity, using the complete basis of gauged curvature-squared supergravity \cite{Ozkan:2013nwa,Gold:2023dfe, Gold:2023ymc,Ozkan:2024euj,Gold:2023ykx}. We show that all three supersymmetric curvature-squared terms contribute to the Euclidean action of the charged rotating AdS black hole regardless of its supersymmetry.

	Before we continue, it is worth mentioning  some closely related work \cite{Bobev:2022bjm, Cassani:2022lrk, Cano:2024tcr, Cassani:2024tvk}. Both \cite{Bobev:2022bjm} and \cite{Cassani:2022lrk} computed 4-derivative corrections to thermodynamics of charged rotating black holes in five dimensional minimal gauged supergravity using only the supersymmetric Weyl-squared and Ricci scalar-squared action. Surprisingly, it is the bare AdS radius rather than the effective AdS radius that enters the BPS relation among conserved charges although the AdS radius and conserved charges are all affected by the curvature squared combinations adopted in \cite{Bobev:2022bjm, Cassani:2022lrk,Cassani:2024tvk}. Thus a third independent check on the thermodynamic quantities is urgently needed. In the very recent work \cite{Cano:2024tcr},  the near horizon geometry of BPS charged rotating black hole was solved perturbatively when the two angular momentum $J_1,\,J_2$ are nearly equal, resulting in the BPS black hole entropy expressed order by order in powers of $J_1-J_2$. Upon taking the BPS limit, our results not only yield the correct black hole entropy in the full parameter region, but also gives rise to a corrected linear relation amongst mass, electric charge and two angular momentum. As the higher derivative corrections to the BPS relation are fully encoded in the effective AdS radius, our result seems more natural from the point of view of AdS superalgebra compared to the previous result \cite{Cassani:2022lrk} which includes only the bare AdS radius.
	%%%%%%%%%%%%%%%%%%%%%%%%%
	%\section{Quadratic curvature corrections to Kerr-AdS  black holes}

	\textit{Quadratic curvature corrections to Kerr-AdS  black holes.}---We consider the effective theory of the Einstein gravity with a negative cosmological constant in five dimensions extended by the general quadratic curvature terms. In Euclidean signature, the action takes the form
	\bea
	I_{QG}&=&-\frac{\sigma_0}{16\pi}\int d^5x\sqrt{g}\left(R+12\ell_0^{-2}+{\cal L}_4\right)\ ,
	\nn\w1
	%%%
	{\cal L}_{4}&=&c_1 R^2 + c_2 r^{\mu\nu} r_{\mu\nu} + c_3 C^{\mu\nu\rho\sigma} C_{\mu\nu\rho\sigma}\ ,
	\label{QG}
	\eea
	where $r_{\mu\nu}=R_{\mu\nu}-\frac{1}{5}g_{\mu\nu}R$, $C_{\m\n\r\s}$ is the Weyl tensor, coefficients $c_i$ are of the dimension length squared and we introduced $\s_0=1/G$ for later convenience. Without higher derivative corrections, the general Kerr-AdS solution has been obtained in \cite{Hawking:1998kw}. Denote
	\be
	\Xi_{a,0}=1-a^2\ell_0^{-2}\ ,\quad \Xi_{b,0}=1-b^2\ell_0^{-2}\ ,
	\ee
	the mass, entropy and two angular momenta are given by \cite{Hawking:1998kw,Gibbons:2004ai}
	\bea
	M_0&=&\frac{\s_0\pi m(2\Xi_{a,0}+2\Xi_{b,0}-\Xi_{a,0}\Xi_{b,0})}{4\Xi_{a,0}^2\Xi_{b,0}^2}\ ,\nn\w1
	%%%
	S_0&=&\frac{\s_0\pi^2(r_0^2+a^2)(r_0^2+b^2)}{2r_0\Xi_a\Xi_b}\ ,\nn\w1
	%%%
	J_{a,0}&=&\frac{\s_0\pi m a}{2\Xi_{a,0}^2\Xi_{b,0}}\ ,\quad J_{b,0}=\frac{\s_0\pi m b}{2\Xi_{b,0}^2\Xi_{a,0}}\ .
	\label{MSJ}
	\eea
	%%%
	where $m,\,a,\,b$ are integration constants and $r_0$ is the radius of the outer horizon determined by
	\be
	(r_0^2+a^2)(r_0^2+b^2)(1+r_0^2\ell_0^{-2})-2m r_0^2=0  \ .
	\label{r0}
	\ee
	Thermodynamic potentials including temperature and two angular velocities are \cite{Hawking:1998kw, Gibbons:2004ai}
	\bea
	T_0&=&\frac{1}{2\pi}\left[\frac{r_0(1+r_0^2\ell_0^{-2})}{r_0^2+a^2}+\frac{r_0(1+r_0^2\ell_0^{-2})}{r_0^2+b^2}
	-\frac{1}{r_0}
	\right]\  ,\nn\w1
	%%%
	\Omega_{a,0}&=&\frac{a(1+r_0^2\ell_0^{-2})}{r_0^2+a^2}\ ,\quad \Omega_{b,0}=\frac{b(1+r_0^2\ell_0^{-2})}{r_0^2+b^2}\ ,
	\label{TO}
	\eea
	Using \eqref{r0}, one can verify that \eqref{MSJ} and \eqref{TO} obey the first law of thermodynamics. The Gibbs free energy is obtained as
	\bea
	G_0&=&M_0-T_0 S_0-\Omega_{a,0} J_{a,0}-\Omega_{b,0} J_{b,0}\nn\w1
	%%%
	&=&\frac{\s_0\pi}{4\Xi_{a,0}\Xi_{b,0}}\left[m-\ell_0^{-2}(r_0^2+a^2)(r_0^2+b^2)\right].
	\eea
	To compute 4-derivative corrections to the thermodynamics of general Kerr-AdS solution,
	we first perform the field redefinitions
	\bea
	g_{\mu\nu}&\rightarrow & g'_{\mu\nu}=g_{\mu\nu}+\lambda_0 g_{\mu\nu}+\lambda_1 R_{\mu\nu}+\lambda_2g_{\mu\nu}R\ ,\nn\w1
	%%%
	\lambda_0&=&\ft{40}{3}c_1\ell_0^{-2}\ ,\quad	\lambda_1=-c_2\ ,\quad  \lambda_2=\ft{2}{3} c_1+\ft{1}{5}c_2\ .
	\label{field redefinition 1}
	\eea
	which transform the general quadratic curvature theory \eqref{QG} to the Einstein-Weyl theory with equivalent thermodynamic variables \cite{Hu:2023gru}
	\bea
	{I}_{EW}&=&-\frac{{\s}}{16\pi}\int d^{5}x\sqrt{g}\left[(R+12{\ell}^{-2})+c_{3}C_{\mu\nu\rho\sigma}C^{\mu\nu\rho\sigma}\right]\nn\w1
	%%%
	&&-I_{\rm surf}\ ,
	\label{QGToWeyl}
	\eea
	where the surface term includes the GHY term and counterterms for the 2-derivative bulk action
	\bea
	I_{\rm surf}&=&\frac{{\s}}{16\pi}\int_{z=\epsilon} d^{4}x\sqrt{h}\big[2K-(\frac{6}{{\ell}}+\frac{\ell}{2}\mathcal{R})+\log\frac{\epsilon^{2}}{\ell^{2}}{\cal A}_4\big],
	\nn\\
	{\cal A}_4&=&\frac{\ell^{3}}{8}(\mathcal{R}^{ij}\mathcal{R}_{ij}-\frac{1}{3}\mathcal{R}^{2})+c_{3}\frac{\ell}{2}\mathcal{C}^{ijkl}\mathcal{C}_{ijkl},
	\label{surf}
	\eea
	where $K$ is the extrinsic curvature of the AdS boundary located at $z=\epsilon$ for $\epsilon\rightarrow 0$, and $\mathcal{R}$, ${\cal R}_{ij}$ and ${\cal C}_{ijkl}$ refer to the boundary Ricci scalar curvature, Ricci tensor and Weyl tensor respectively. The logarithmic terms  induced by the Einstein gravity were well known \cite{Henningson:1998gx}. However, we also find that the bulk  Weyl tensor squared also induces a new logarithmic counterterm proportional to $c_3$. As in the case of 2-derivative pure gravity \cite{Papadimitriou:2004ap}, these surface terms are sufficient to remove all the IR divergences for general AlAdS solutions. Notice that the new logarithmic counterterm proportional to $c_3$ is absent when the bulk spacetime dimension is even (see \cite{Grumiller:2013mxa} for $D=4$ case and \cite{Anastasiou:2020mik, Anastasiou:2023oro} for $D=6$ case). Moreover, all the logarithmic terms vanish for AdS black holes with $S^1\times M_3$ type boundary topology for $M_3$ being Einstein which is the case for rotating Kerr-AdS black hole. The coefficients of the logarithmic counterterms also imply the central charges in the dual CFT \cite{Fukuma:2001uf}
\be
{\rm a}=\frac{\pi{\ell}^3}{8}{\sigma},\quad {\rm c}=\frac{\pi\ell^{3}}{8}\sigma(1+8c_{3}\ell^{-2})\ .
\label{ac}
\ee

By treating the 4-derivative terms perturbatively, {\it i.e.}, their corrections to the solutions vanish smoothly as the 4-derivative couplings are turned off,  one can still impose the standard Dirichlet boundary condition on the metric residing on the conformal boundary \cite{Papadimitriou:2004ap,Papadimitriou:2005ii,Anastasiou:2020zwc}.

In \eqref{QGToWeyl} and \eqref{surf}, various coupling constants in the Einstein-Weyl theory are related to the original ones in \eqref{QG} by
	%%%%
	\be
	{ \sigma}=\sigma_0 -\frac{40 c_1 \sigma_0 }{\ell _0^2}\ ,\quad
	{\ell}=\ell_0-\frac{10 c_1}{3 \ell _0}\ .
	\label{corel}
	\ee
	%%%%	
	For Einstein-Weyl gravity, the quadratic curvature correction to the Euclidean action of Kerr-AdS  black hole is obtained by simply plugging the uncorrected solution \cite{Hawking:1998kw} into Weyl-squared action and performing the integration \cite{Hu:2023gru}. With the Dirichlet boundary condition, the resulting Euclidean action is defined in the grand ensemble with fixed temperature and angular velocity. In terms of these variables, the higher derivative interactions only affect the form of the functional dependence of the Euclidean action on $T$ and $\O_{a,b}$. Schematically we have
	\be
	I_{EW}=I_0(T,\O)+c_3 I_1(T,\O)+\frac{3{\rm a}}{4\ell}\ .
	\ee
	 According to the prescription of \cite{Reall:2019sah, Hu:2023gru}, in terms of the variables $r_0,\,a,\, b$, the form of temperature and angualr momenta remain the same, namely
	\be
	T=T_0\ ,\quad \Omega_a=\Omega_{a,0}\ ,\quad \Omega_b=\Omega_{b,0} \ ,
	\ee
	whereas the conserved charges do receive explicit higher derivative corrections.
	
	Expressing parameters of Einstein-Weyl theory back to those of the original theory, we obtain the Euclidean action of Kerr-AdS  black hole with general quadratic curvature corrections. Using the standard relation between Euclidean action and Gibbs free energy \cite{Gibbons:1976ue} and omitting the central charge , we obtain Gibbs free energy of Kerr-AdS  black hole in the grand canonical ensemble
	%%%
	\be
	G_{QG}=G_0+\delta G_1+\delta G_2 \ ,
	\label{FreeEnergyQG}
	\ee
	%%%
	in which $\d G_1$ and $\d G_2$ are corrections caused by the Ricci scalar squared and Weyl-squared
	%%%
	\begin{widetext}
		\bea
		\delta G_1&=&-c_1 \frac{5\pi(\hat{a}^{2}+1)(\hat{b}^{2}+1) \hat{r}_{0}^{2}\sigma_0}{6\hat{\Xi}_{a,0}^{2}\hat{\Xi}_{b,0}^{2}}\left[7\hat{r}_{0}^{6}\hat{a}^{2}\hat{b}^{2}-8\hat{r}_{0}^{4}\big(\hat{a}^{2}\hat{b}^{2}+\hat{a}^2+\hat{b}^{2}\big)+\hat{r}_{0}^{2}(7\hat{a}^{2}+7\hat{b}^{2}+9)-6\right]\ ,\cr
		%%%
		\delta G_2&=&
		-c_{3}\frac{\pi (\hat{r}_{0}^{2}+1){}^{2}\sigma_0\big[\hat{a}^{2}\hat{b}^{2}(\hat{a}^{2}\hat{b}^{2}-20)+2(\hat{a}^{2}+\hat{b}^{2})(1-3\hat{a}^{2}\hat{b}^{2})+\hat{a}^{4}+\hat{b}^{4}+9\big]}{4(\hat{a}^{2}+1)(\hat{b}^{2}+1)\hat{\Xi}_{a,0}\hat{\Xi}_{b,0}}\ .\label{free energy}
		\eea
	\end{widetext}
	%%%%%%%
	where we introduced dimensionless variables $\hat{r}_{0}=r_{0}\ell_{0}^{-1},\ \hat{a}=a{r}_{0}^{-1},\ \hat{b}=b{r}_{0}^{-1}$ and accordingly $\hat{\Xi}_{a,0}=1-\hat{a}^{2}\hat{r}_{0}^{2},\ \hat{\Xi}_{b,0}=1-\hat{b}^{2}\hat{r}_{0}^{2}$. It is interesting to see that $r_{\m\n}r^{\m\n}$ does not contribute. The other thermodynamic variables are obtained from the Gibbs free energy via standard relations
	\bea
	M&=& G_{QG}+TS+\O_{a}J_{a}+\O_{b}J_{b}\ ,\nn\w1
	%%%
	S&=&-\frac{\partial G_{QG}}{\partial T}\Big{\vert}_{\O_{a,b}},\ ~ J_{a(b)}=-\frac{\partial G_{QG}}{\partial \O_{a(b)}}\Big{\vert}_{T,\O_{b(a)}} .
	\eea
	To test results above, we consider Kerr-AdS  black hole with equal rotation which is cohomogenity-1 allowing us to solve the 4-derivative field equations numerically.  From the numerical solution, we extract the mass and angular momenta using the generalized AMD formula for quadratic curvature theories \cite{Ashtekar:1999jx,Okuyama:2005fg,Pang:2011cs} and compare them to the analytical results obtained above. For non-extremal Kerr-AdS solution, the comparison is still not easy as all the thermodynamic quantities depend on two variables. To simplify the comparison further, we restrict to the extremal Kerr-AdS  solution in which all the thermodynamic quantities depend on a single variable. In particular, we consider mass as a function of the angular momentum
	\be
	M_{\rm ex}(J_{\rm ex})=M_0(J_{\rm ex})+\delta M(J_{\rm ex})\ ,
	\ee
	of which $M_0(J_{\rm ex})$ can be derived from the uncorrected quantities given in \eqref{MSJ}, the explicit expression of $M_{\rm ex}(J_{\rm ex})$ is a bit lengthy and will be postponed to the supplemental material. It is the latter which will be subject to numerical tests.
	%%%%%%%%%%%%%%%%%%%%%%%%%%%%%

	In Fig \ref{dMJ} below, we show that the analytical and numerical results indeed match in a wide range of variables for the well studied Gauss-Bonnet (GB) combination corresponding to $c_1=\frac{3}{10}\alpha,\ c_2=-\frac{8}{3}\alpha,\  c_3=\alpha$ and Einstein-Weyl (EW) gravity corresponding to $c_1=c_2=0,\ c_3=\alpha$. In the plot, we have defined the dimensionless mass $\mathcal{M}=M_{{\rm ex}}/(\pi\sigma_0\ell_{0}^{2})$ and angular $\mathcal{J}=J_{{\rm ex}}/(\pi\sigma_0\ell_{0}^{3})$.
The plot is independent of choice of $\a$ as $\delta{\cal M}$ depends on $\a$ linearly.
%%%%%%%%%%%%%%
	\begin{figure}[ht]
		\centering
		\includegraphics[scale=0.85]{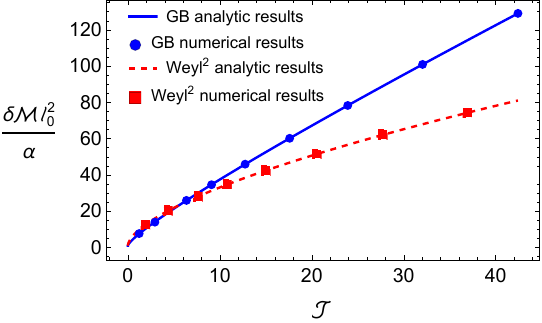}
		\caption{For extremal Kerr-AdS  black hole with equal rotation, we exhibit matching between analytical and numerical results for 5$D$ Einstein-Gauss-Bonnet gravity and Einstein Weyl gravity. }
		\label{dMJ}
	\end{figure}
	%%%%%%%%%%%%%%%%%%%%%%%%%

	%\section{Charged rotating black holes in $D=5$  minimal gauged supergravity with complete $R^2$ corrections }
	
	\textit{Quadratic curvature corrections to charged rotating  black holes in $5D$ minimal gauged supergravity.}---The complete basis of curvature squared supergravity in $5D$ minimal gauged supergravity
	was presented in \cite{Gold:2023ymc,Ozkan:2024euj}. After applying field redefinitions preserving black hole thermodynamics \cite{Hu:2023gru}, we obtain the action below
	
	\bea
	&&I_{5D,N=1}=-\frac{\sigma}{16\pi}\int d^{5}x\sqrt{g}\big(R+\frac{12}{{\ell}^{2}}-\frac{1}{4{g}^{2}}F_{\m\n}F^{\m\n} \nn\w1
	%%%
	&& +\frac{\rm i}{12\sqrt{3}{g}^{3}}\epsilon^{\mu\nu\rho\sigma\delta}F_{\mu\nu}F_{\rho\sigma}A_{\delta}+c_1{\mathcal{L}}_{{\rm Weyl}^2}\big) -I_{\rm surf}\,,
	\label{5dsugra}
	\eea
	%%%
	where the surface term includes addiditional logarithmic counterterm proportional to $F_{ij}F^{ij}$ \cite{Taylor:2000xw}.  We recall that in the original Lagrangian, there are supersymmetric Ricci tensor squared and Ricci scalar squared actions. Denoting their coefficients by $c_2$ and $c_3$ respectively, the coupling constants in the action \eqref{5dsugra} are related to the original ones via (See supplementary material for further details)
	%%%$
	\bea
	{\sigma}&=&\sigma_0-24\s_0(c_{2}+c_{3})\ell_{0}^{-2},\ \  {g}=g_0\big(1+4(c_{2}+c_{3})\ell_{0}^{-2}\big),
	\nn\w2
	{\ell}&=&\ell_{0}\big(1-4(c_{2}+c_{3})\ell_{0}^{-2}\big)\ ,
	\eea
	%%%
	where $g_0$ is the U(1) coupling before the field redefinition. It is a bookkeeping parameter introduced via $A_\m\rightarrow A_\m/g_0$ in the standard supergravity action and does not affect physical quantities.
	The on-shell Weyl-squared supergravity action  ${\mathcal{L}}_{{\rm Weyl}^2}$ is given by
	\bea
	&&{\mathcal{L}}_{{\rm Weyl}^2}=-\frac{2}{{g}^{2}{\ell}
		^2}F_{\m\n}F^{\m\n}-\frac{\rm i}{\sqrt{3}{g}^{3}{\ell}^2}\epsilon^{\mu\nu\rho\sigma\delta}F_{\mu\nu}F_{\rho\sigma}A_{\delta}\cr
	%%%
	&&+C_{\mu\nu\rho\sigma}C^{\mu\nu\rho\sigma}-\frac{1}{2{g}^{2}}C_{\mu\nu\rho\sigma}F^{\mu\nu}F^{\rho\sigma} +\frac{13}{96{g}^{4}}(F_{\m\n}F^{\m\n})^{2}\cr
	&&-\frac{13}{24{g}^{4}}F^4
	+\frac{\sqrt{3}\rm i}{6{g}}\epsilon^{\mu\nu\rho\sigma\alpha}A_{\mu}C_{\nu\rho}^{\ \ \ \beta\gamma}C_{\sigma\alpha\beta\gamma}\ ,
	\eea
	where we define $F^4:=F_{\m\n}F^{\n\l}F_{\l\d}F^{\d\m}$. Different from the ungauged supergravity, in the gauged case, the curvature squared supergravity actions contain also
	2-derivative terms.  It is worth mentioning that in the frame we have chosen, the parameter $\ell$ already represents the effective AdS radius. This is quite different from the setup adopted in \cite{Bobev:2022bjm, Cassani:2022lrk} where the curvature squared combinations do renormalize the bare AdS radius \footnote{When the curvature squared combinations contribute to the effective AdS radius, computing corrections to the Euclidean action using the original method of \cite{Reall:2019sah} must be excercised with caution. The correct expression for temperature, electric potential and angular velocities involve an extra factor $\g=\ell_0/\ell_{\rm eff}$. See \cite{Hu:2023gru} for example in the case of Einstein-Gauss-Bonnet gravity. Previous work \cite{Cassani:2022lrk} seems to have overlooked this factor although in their case, $\g\neq 1$. Consequently, the conserved charges obtainted by differentiating Euclidean action with respect to these variable are incorrect.}.
	
	The general charged rotating black hole solution in 2-derivative $5D$ minimal gauged supergravity was obtained in \cite{Chong:2005hr}. Regularity of the solution determines the inverse temperature to be
	%%%%%%%%%%%
	\be
	\b=\frac{2\pi r_{0}\left(abq+(r_{0}^{2}+a^{2})(r_{0}^{2}+b^{2})\right)}{r_{0}^{4}\left[1+(a^{2}+b^{2}+2r_{0}^{2}){\ell}^{-2}\right]-(ab+q)^{2}}\ ,
	\label{Tem}
	\ee
	%%%%%%%%%%%
	where $r_0$ is the radius of the outer horizon, $a,\, b,\,q$ are parameters related to angular velocities and electrostatic potential given by
	\bea
	&&{\Omega}_{a}=\frac{a(r_{0}^{2}+b^{2})(1+{\ell}^{-2}r_{0}^{2})+bq}{abq+(r_{0}^{2}+a^{2})(r_{0}^{2}+b^{2})}\ ,\nn\w1
	&&{\Omega}_{b}=\frac{b(r_{0}^{2}+a^{2})(1+{\ell}^{-2}r_{0}^{2})+aq}{abq+(r_{0}^{2}+a^{2})(r_{0}^{2}+b^{2})}\ , \nn\w1
	&&{\Phi}_{e}=\frac{{g}\sqrt{3}qr_{0}^{2}}{abq+(r_{0}^{2}+a^{2})(r_{0}^{2}+b^{2})}\ .
	%%%
	\label{tp}
	\eea
	%%%%%%%%%%%
	Analogous to the discussion in the previous section, these variables characterize the Euclidean action of charged rotating black holes as the current choice of boundary condition specifies the grand canonical ensemble.
	
	With supersymmetric curvature squared corrections switched on, in principle one would have to first solve for the modified field equations before computing the corrected on-shell action. However, as established in \cite{Reall:2019sah,Hu:2023gru} and here, this step can be circumvented if the higher curvature terms are expressed in terms of the Weyl tensor. The fully corrected on-shell action can be obtained by simply evaluating the modified action on the uncorrected solution. Adopting this method, we obtain the Euclidean action for the general charged rotating AdS black holes in $5D$ minimal gauged supergravity extended by all three curvature squared invariants. The full result is given in the supplemental material. Similar to the non-supersymmetric case, the thermodynamic variables $\b,\,\O_{a,b},\, \Phi_e$ remains the same form as in \eqref{tp}. Thus the conserved charges can be obtained by differentiating the Euclidean action with respect to $\b,\,\O_{a,b},\, \Phi_e$.
	
	We now apply the result to under the entropy of the supersymmetric charged rotating AdS black hole \cite{Gutowski:2004ez,Gutowski:2004yv} which admits microscopic description in terms of index of the dual $4D,\,N=1$ superconformal field theory \cite{Cabo-Bizet:2018ehj, Choi:2018hmj,Benini:2018ywd, Honda:2019cio, David:2020ems, Agarwal:2020zwm, Benini:2020gjh}. To proceed, we impose supersymmetry condition
	\be
	q=-(a-ir_{0})(b-ir_{0})(1-ir_{0}{\ell}^{-1})\ .
	\label{susy}
	\ee
	Note that the BPS limit requires also zero temperature and can be arrived via $r_{0}\rightarrow\sqrt{{\ell}(a+b)+ab}$ \cite{Chong:2005hr}. Different from previous work \cite{Bobev:2022bjm,Cassani:2022lrk}, the supersymmetric condition now is corrected by the 4-derivative terms whose effect is fully encoded in the effective AdS radius $\ell$. Subsequently, one can define thermodynamic potentials
	\bea
	\omega_{a}&=&{\beta}_{s}({\Omega}_{a,s}-{\Omega}_{a,*})=\frac{2\pi(b-ir_{0})(a-{\ell})}{\Xi}\ ,\nn \\
	%%%
	\omega_{b}&=&{\beta}_{s}({\Omega}_{b,s}-{\Omega}_{b,*})=\frac{2\pi(a-ir_{0})(b-{\ell})}{\Xi}\ ,\nn \\
	%%%
	\varphi &=&{\beta}_{s}({\Phi}_{e,s}-{\Phi}_{e,*})=\frac{6{\pi g\ell  }(a-ir_{0})(b-ir_{0})}{\sqrt{3}\Xi},
	\eea
	%%%%%%%%%%%%%%%%%%%%%%%%%%
	where $\Xi= 2r_{0}({\ell}+a+b)+i\big({\ell}(a+b)+ab\big)-3ir_{0}^{2}$ and they
	satisfy $
	\omega_{a}+\omega_{b}-\frac{\sqrt{3}}{g{\ell}}\varphi-2\pi i=0$.
	%%%
	Here ``$s$" means the supersymmetry condition \eqref{susy} has been applied and ``*" denotes values of these variable in the BPS limit
	\be
	{\Omega}_{a,*}={\ell}^{-1}\ ,\quad {\Omega}_{b,*}={\ell}^{-1}\ ,\quad {\Phi}_{e,*}=\sqrt{3}g\ ,
	\ee
	Imposing the supersymmetric condition \eqref{susy}, we find that the Euclidean action drastically simplifies
	%%%%
	\be
	I_{{\rm ren},s}=\frac{\pi\sigma\varphi^{3}(1-\frac{12c_{1}}{\ell^{2}})}{12\sqrt{3}g^{3}\omega_{a}\omega_{b}}+\frac{c_{1}\pi\sigma\varphi(\omega_{a}^{2}+\omega_{b}^{2}-4\pi^{2})}{\sqrt{3}g\omega_{a}\omega_{b}}\ .
	\label{regI}
	\ee
	In terms of the ${\rm a}$ and ${\rm c}$ central charges of the dual theory given in \eqref{ac} with $c_3$ replaced by $c_1$,
	the supersymmetric on-shell action indeed takes the form as its counterpart in the dual field theory \cite{Cassani:2021fyv}.  Although our supersymmetric action takes the same form as previous results in \cite{Bobev:2022bjm,Cassani:2022lrk}, the details are different
	as our $\omega_{a,b},\varphi$ defined in (24) depend on the effective AdS radius, while those in \cite{Bobev:2022bjm,Cassani:2022lrk} use bare AdS radius instead. Taking the BPS limit, we find the conserved charges obey the linear relation \footnote{A similar equality was obtained in \cite{Cassani:2022lrk} where it is the bare AdS radius rather than the effective AdS radius that enters the expression. }
	%%%%%%
	\be
	M_{*}-{\ell}^{-1}J_{a,*}-{\ell}^{-1}J_{b,*}-\frac{3}{2}{\ell}^{-1}Q_{R}=0\ ,
	\ee
	where $Q_{R}:=\frac{2g{\ell}}{\sqrt{3}}Q_{e,*}$ is the canonically normalized U(1) R-charge in the dual SCFT. This equality leads to vanishing Gibbs free energy.
	In BPS limit, the entropy of the charged rotating black hole also reproduces the microscopic result \cite{Bobev:2022bjm, Cassani:2022lrk}. Namely, up to ${\cal O}(c_i)$ it's given by
	\begin{widetext}
		\be
		S_{*}=\pi\sqrt{3Q_{R}^{2}-8\mathrm{a}\left(J_{a,*}+J_{b,*}\right)
			-16\mathrm{a}(\mathrm{a}-\mathrm{c})\frac{\left(J_{a,*}
				-J_{b,*}\right){}^{2}}{Q_{R}^{2}-2\mathrm{a}\left(J_{a,*}+J_{b,*}\right)}}\ .
		\ee
	\end{widetext}
	
	\textit{Conclusion and outlook.}---So far we have showed that Weyl gravity offers an efficient way of computing the leading higher curvature contributions to thermodynamic quantities of general rotating AdS black holes in $5D$ quadratic gravity theories with or without supersymmetry. In fact, this approach also applies to more general quadratic gravity theories with matter couplings in other dimensions simply due to the fact that Weyl tensor vanishes sufficiently fast near the AdS boundary.  We believe our approach can be pushed forward to the next to next to leading order higher curvature corrections as already achieved in the asymptotically flat case \cite{Ma:2023qqj}. Together with the first law of thermodynamics,  our results also imply that in the basis of Weyl tensor the leading higher derivative corrections to the black entropy can be readily computed via $\delta S=-I_{\rm hd}|_{(T,\Phi_e,\O_{a,b})}$  which should be useful in the discussion of the AdS counterpart of weak gravity conjecture. To further establish the effectiveness of Weyl gravity in AdS quantum gravity through holography, it should be very interesting to consider other solutions such AdS black strings \cite{Hosseini:2019lkt,Hosseini:2020vgl,Bobev:2021qxx} and compute correlation functions as well as various transport coefficients.
	
	\textit{Acknowledgement.}---We are grateful to D.~Cassani, M.~Ozkan, A.~Ruiprez, G.~Tartaglino-Mazzucchelli and E.~Turetta for useful discussions. In particular, we thank D.~Cassani, A.~Ruiprez and E.~Turetta
	for communications regarding the dependence of the onshell action on the combination of $c_2+c_3$.
	We also thank N. Bobev for useful correspondence. H.L.~is supported in part by the National Natural Science Foundation of China (NSFC) grants No.~11935009 and No.~12375052. Y.P.~is supported by the National Key Research and Development Program No.~2022YFE0134300 and the NSFC grant No.~12175164.

\clearpage
\widetext
\section{Supplementary Material}
\subsection{A. Numerical analysis}

Kerr-AdS black holes in general dimensions were constructed in \cite{Gibbons:2004uw,Gibbons:2004js}. In particular, the general Kerr-AdS metric with two independent angular momenta in 5$D$ were given in \cite{Hawking:1998kw}. When the two angular momenta are equal, i.e.,$a=b$, the 5$D$ Kerr-AdS metric can be written as \cite{Feng:2016dbw}
\bea \label{ds2b=a}
ds^2&=&-\frac{h}{W}dt^2+\frac{d\rho^2}{f}+\frac{\rho^2}{4}W(\sigma_3+\omega dt)^2+\frac{\rho^2}{4}d\Omega_2^2\ ,\qquad W_0=1+\frac{\nu^2}{\rho^4},\qquad \omega_0=\frac{2\sqrt{2\mu}\nu}{\rho^4+\nu^2}\ ,\cr
%%%
d\Omega_2^2&=&d\theta^2+\sin^2\theta d\varphi^2\ ,\quad \sigma_3=d\psi+\cos\theta d\varphi\ , \qquad	h_0=f_0=\left(1+\ell_0^{-2} \rho ^2\right) \left(1+\frac{\nu ^2}{\rho ^4}\right)-\frac{2 \mu }{\rho ^2}\ .
\eea
The level surfaces are the squashed 3-spheres written as a $U(1)$ bundle over $S^2$.
The metric is asymptotic to AdS$_5$, with mass $M_0$ and (two equal) angular momentum $J_0$:
\be
M_0=\frac{\sigma_0\,\pi}{8}(6\mu + \ell_0^{-2} \nu^2)\ ,\qquad J_0= \frac{\sigma_0\,\pi}{4}\, \sqrt{2\mu} \nu\ .
\ee
The metric describes a black hole for appropriate $(M_0, J_0)$ when $h_0=f_0$ has a real root. The black hole becomes extremal when we have a double root $\rho_0$, corresponding to taking
\be
\mu=\rho_0^2 (1 + \ell_0^{-2} \rho_0^2)^2\ ,\qquad \nu= \rho_0^2 \sqrt{1 + 2 \ell_0^{-2} \rho_0^2} \ .\label{extremal}
\ee
The perturbative ansatz of metric functions for the general quadratic gravity \eqref{QG} in coordinate \eqref{ds2b=a} are
\bea
h=h_0(1+\delta h)\ ,\quad f=f_0(1+\delta f)\ ,\quad W=W_0+\delta W\ ,\quad \omega=\omega_0+\delta\omega\ .
\eea
At large $\rho$, $\delta f$ takes the form
\bea
\delta f&=&\frac{20c_1}{3}\Big(\frac1{\ell_0^2} - \frac{1}{\rho^2}\Big)+\frac{f_{i,4}}{\rho^4}+
\frac{f_{i,6}}{\rho^6}+\frac{f_{i,8}}{\rho^8}+\frac{f_{i,10}}{\rho^{10}}
+ \cdots +f_l\log\rho\left(\frac{1}{\rho^4}+\frac{f_{i,l,6}}{\rho^6}
+\frac{f_{i,l,8}}{\rho^8}+\frac{f_{i,l,10}}{\rho^{10}} + \cdots\right).\label{solution f at infinity}
\eea
where the ellipses denote the higher-order terms in the large $\rho$ power expansion.  The four leading-term coefficients $f_l$,  $f_{i,4}$, $f_{i,6}$, $f_{i,8}$ are free parameters, with the rest of the coefficients solved in terms of these four, by the equations of motion order by order. However, the coefficient $f_l$ must be zero to avoid logarithmic divergence of large $\rho$. Near the horizon, we have

\bea
\delta f&=&f_{h,0}+f_{h,1}(\rho-\rho_0)+f_{h,2}(\rho-\rho_0)^2+f_{h,3}(\rho-\rho_0)^3+f_{h,4}(\rho-\rho_0)^4\cr
%%%
&&+f_{i,h,0}(\rho-\rho_0)^{-\frac{3}{2}+\frac{ \sqrt{5}}{2} \sqrt{\frac{7 \rho_0^2\ell_0^{-2}+5}{3 \rho_0^2\ell_0^{-2}+1}}}
\Big[1+f_{i,h,1}(\rho-\rho_0)+f_{i,h,2}(\rho-\rho_0)^2\cr
%%%
&&
+f_{i,h,3}(\rho-\rho_0)^3+f_{i,h,4}(\rho-\rho_0)^4\Big]\ .\label{solution f at horizon}
\eea
In this case, there is only one free parameter $f_{i,h,0}$ for fixed $\rho_0$, with the remaining coefficients all solved in terms of $f_{i,h,0}$ by equations of motion order by order. The global geometric implication of the irrational power in the near-horizon power-series expansion was studied in \cite{Mao:2023qxq} for the asymptotically-flat examples.

To perform numerical calculation, we can choose a fixed $\rho_0=1$, and then select a value for $f_{i,h,0}$ and integrate it from $\rho_0$ to a large $\rho \sim 200$ and perform data fitting with \eqref{solution f at infinity} and read off the free for free parameters $f_{i,4}$, $f_{i,6}$, $f_{i,8}$. For a generic value of $f_{i,h,0}$, $f_l$ is nonzero. We thus need to fine-tune  the parameter $f_{i,h,0}$ so that $f_l=0$. For each given value of $\rho_0$, there is only one such $f_{i,h,0}$, which implies that the remainder of the three free parameters $f_{i,4}$, $f_{i,6}$, $f_{i,8}$ are now all fixed as a function of $\rho_0$. The full leading-order perturbed solution thus has only one parameter, namely the radial location of the horizon.

For quadratic theory \eqref{QG}, the AMD formula is \cite{Pang:2011cs,Ashtekar:1999jx,Okuyama:2005fg}
\bea\label{AMD}
Q[K]=\frac{\sigma_0\ell\Xi_{\rm AMD}}{8\pi(D-3)}\oint d\bar{\Sigma}_\mu\bar{\mathcal{E}}^\mu_{\ \nu}K^\nu\ ,
\eea
where $D=5$ and
\be
\Xi_{\rm AMD}=1-\frac{2}{\ell^2}\big(c_1 (D-1) D-2 c_3 (D-3)\big)\ ,\quad \ell=\ell _0-\frac{c_1 (D-4) (D-1) D}{2 (D-2) \ell _0}\ .
%%%
\ee

The mass and angular momentum under the perturbation in the numerical approach can be first obtained by using AMD formula \eqref{AMD} in terms of $(\mu,\nu)$ and the parameters of the perturbed solution at large $\rho$ expansion. We then take the  extremal limit \eqref{extremal} and we have
\bea \label{Num}
\mathcal{M}_{\rm Num}&=&\frac{Q[\partial_t]}{\pi \sigma_0\ell_0^2}=\frac{3\hat{\rho}_{0}^{2}}{4}+
\frac{13\hat{\rho}_{0}^{4}}{8}+\hat{\rho}_{0}^{6}+
\frac{1}{8}(\hat{f}_{i,4}+4\hat{f}_{i,6})
-\frac{5}{6}\hat{c}_{1}(-3+22\hat{\rho}_{0}^{2}+49\hat{\rho}_{0}^{4}
+32\hat{\rho}_{0}^{6})+\hat{c}_{3}(6+13\hat{\rho}_{0}^{2}
+8\hat{\rho}_{0}^{4})\hat{\rho}_{0}^{2}\ ,\cr
%%%
\mathcal{J}_{\rm Num}&=&\frac{Q[\partial_\psi]}{\pi \sigma_0\ell_0^3}=		\frac{\hat{\rho}_{0}^{3}}{2\sqrt{2}}\sqrt{1+2\hat{\rho}_{0}^{2}}(1+\hat{\rho}_{0}^{2})
+\frac{3(4\hat{\rho}_{0}^{4}+17\hat{\rho}_{0}^{2}+10)\hat{\rho}_{0}^{2}
	\hat{f}_{i,4}+3(8\hat{\rho}_{0}^{6}+10\hat{\rho}_{0}^{4}+4\hat{\rho}_{0}^{2}-3)
	\hat{f}_{i,6}-9\hat{f}_{i,8}}{48\sqrt{2}\hat{\rho}_{0}^{3}(1+\hat{\rho}_{0}^{2})
	\sqrt{1+2\hat{\rho}_{0}^{2}}}\cr	
%%%
&& +\frac{\hat{\rho}_{0}\big(-10\hat{c}_{1}(72\hat{\rho}_{0}^{8}+168\hat{\rho}_{0}^{6}
	+112\hat{\rho}_{0}^{4}-2\hat{\rho}_{0}^{2}-13)+3\hat{c}_{3}(64\hat{\rho}_{0}^{8}
	+160\hat{\rho}_{0}^{6}+155\hat{\rho}_{0}^{4}+68\hat{\rho}_{0}^{2}+12)\big)
}{24\sqrt{2}(1+\hat{\rho}_{0}^{2})\sqrt{1+2\hat{\rho}_{0}^{2}}}\ ,
\eea
where $\hat{\rho}_{0}=\rho_0 \ell_0^{-1}$,\ $\mathcal{M}=\frac{M_{{\rm ex}}}{\pi\sigma_0\ell_{0}^{2}},\ \mathcal{J}=\frac{J_{{\rm ex}}}{\pi\sigma_0\ell_{0}^{3}},\ \hat{c}_i=c_i\ell_0^{-2},\ \hat{f}_{i,n}=f_{i,n}\ell_{0}^{-n}$.	

From the free energy \eqref{FreeEnergyQG}, we can compute  the  analytic  (RS) \cite{Reall:2019sah,Hu:2023gru} perturbed thermodynamic quantities. In order to compare these analytic results with the numerical ones \eqref{Num}, we need to perform the coordinate transformation $\hat{r}_0=\frac{\hat{\rho}_0}{\sqrt{2} \sqrt{1+\hat{\rho}_0^2}}$ so that their metrics match. We therefore have
\bea
\mathcal{M}_{\rm RS}&=&\frac{3\hat{\rho}_{0}^{2}}{4}+\frac{13\hat{\rho}_{0}^{4}}{8}
+\hat{\rho}_{0}^{6}+\frac{5}{6}\hat{c}_{1}(-36-53\hat{\rho}_{0}^{2}
+20\hat{\rho}_{0}^{4}+48\hat{\rho}_{0}^{6})\hat{\rho}_{0}^{2}
+\hat{c}_{3}(1+15\hat{\rho}_{0}^{2}+\frac{155\hat{\rho}_{0}^{4}}{4}
+30\hat{\rho}_{0}^{6})\ ,\cr
%%%
\mathcal{J}_{\rm RS}&=&\frac{\hat{\rho}_{0}^{3}}{2\sqrt{2}}\sqrt{1+2\hat{\rho}_{0}^{2}}
(1+\hat{\rho}_{0}^{2})+\frac{5\hat{\rho}_{0}^{3}}{3\sqrt{2}}
\sqrt{1+2\hat{\rho}_{0}^{2}}\big(2\hat{c}_{1}(-6-\hat{\rho}_{0}^{2}
+6\hat{\rho}_{0}^{4})+3\hat{c}_{3}(2+3\hat{\rho}_{0}^{2})\big)\ .
%%%
\eea
Furthermore, we should redefine $\hat{\rho}_{0}$ to fix the angular momentum $\mathcal{J}_0=\frac{\hat{\rho}_{0}^{3}}{2\sqrt{2}}\sqrt{1+2\hat{\rho}_{0}^{2}}
(1+\hat{\rho}_{0}^{2})$. For $\mathcal{M}_{\rm Num},\ \mathcal{J}_{\rm Num}$, we take
\bea
\hat{\rho}_{0}&\to&\hat{\rho}_{0,\rm Num}=\hat{\rho}_{0}+\delta_{\rm Num}\hat{\rho}_{0}\ ,\cr
%%%
\text{\ensuremath{\delta_{\rm Num}\hat{\rho}_{0}}}&=&\frac{1}{24\hat{\rho}_{0}^{5}(1+\hat{\rho}_{0}^{2})
	(1+3\hat{\rho}_{0}^{2})(3+4\hat{\rho}_{0}^{2})}\big[-3(10+17\hat{\rho}_{0}^{2}
+4\hat{\rho}_{0}^{4})\hat{\rho}_{0}^{2}\hat{f}_{i,4}+3(3-4\hat{\rho}_{0}^{2}
-10\hat{\rho}_{0}^{4}-8\hat{\rho}_{0}^{6})\hat{f}_{i,6}+9\hat{f}_{i,8} \cr
%%%
&&+2\hat{\rho}_{0}^{4}\big(10\hat{c}_{1}(-13-2\hat{\rho}_{0}^{2}
+112\hat{\rho}_{0}^{4}+168\hat{\rho}_{0}^{6}+72\hat{\rho}_{0}^{8})
-3\hat{c}_{3}(12+68\hat{\rho}_{0}^{2}+155\hat{\rho}_{0}^{4}
+160\hat{\rho}_{0}^{6}+64\hat{\rho}_{0}^{8})\big)\big]\ .
\eea
For $\mathcal{M}_{\rm RS},\ \mathcal{J}_{\rm RS}$, we take
\bea
\hat{\rho}_{0}&\to&\hat{\rho}_{0,\rm RS}=\hat{\rho}_{0}+\delta_{\rm RS}\hat{\rho}_{0}\ ,\cr
%%%
\text{\ensuremath{\delta_{\rm RS}\hat{\rho}_{0}}}&=&-\frac{10\hat{\rho}_{0}(1+2\hat{\rho}_{0}^{2})
	\big(3\hat{c}_{3}(2+3\hat{\rho}_{0}^{2})+2\hat{c}_{1}(-6-\hat{\rho}_{0}^{2}
	+6\hat{\rho}_{0}^{4})\big)}{9+39\hat{\rho}_{0}^{2}+36\hat{\rho}_{0}^{4}}\ .
\eea
Finally, we obtain the extremal mass with the fixed angular
momentum $\mathcal{J}_{0}$ for both numerical and RS approaches:
\bea
\delta\mathcal{M}_{\text{Num}}&=&\frac{1}{48\hat{\rho}_{0}^{4}
	(1+\hat{\rho}_{0}^{2})}\big[-3(10+15\hat{\rho}_{0}^{2}+2\hat{\rho}_{0}^{4})
\hat{\rho}_{0}^{2}\hat{f}_{i,4}-3(-3+4\hat{\rho}_{0}^{2}+2\hat{\rho}_{0}^{4})
\hat{f}_{i,6}+9\hat{f}_{i,8} \cr
%%%
&&+2\hat{\rho}_{0}^{4}\big(3\hat{c}_{3}(-12-20\hat{\rho}_{0}^{2}-3\hat{\rho}_{0}^{4}
+8\hat{\rho}_{0}^{6})+10\hat{c}_{1}(-7-40\hat{\rho}_{0}^{2}-30\hat{\rho}_{0}^{4}
+6\hat{\rho}_{0}^{6}+8\hat{\rho}_{0}^{8})\big)\big]\ , \cr
\delta\mathcal{M}_{\text{RS}}&=&\frac{5}{6}\hat{c}_{1}
(-12-\hat{\rho}_{0}^{2}+4\hat{\rho}_{0}^{4})\hat{\rho}_{0}^{2}
+\hat{c}_{3}(1+5\hat{\rho}_{0}^{2}+\frac{15\hat{\rho}_{0}^{4}}{4})\ .
\eea
For the numerical approach, we have shown that the coefficients $f_{i,4}$, $f_{i,6}$ and $f_{i,8}$ are all functions of $\rho_0$ only; therefore, both $\delta\mathcal{M}_{\text{Num}}$ and $\delta\mathcal{M}_{\text{RS}}$ are functions of $\rho_0$ only. The Fig.\ref{dMJ} illustrates they match perfectly.

\subsection{B. The 4-derivative effective action}
We start by considering the 4-derivative extension of minimal gauged supergravity in
$D=5$ \cite{Gold:2023ymc,Ozkan:2024euj}
\begin{align}
	I_{5D,\:N=1} & =\frac{\sigma_{0}}{16\pi}\int\sqrt{-g}d^{5}x(\mathcal{L}_{0}+c_{1}\mathcal{L}_{{\rm Weyl}^{2}}+c_{2}\mathcal{L}_{{\rm Ricci}^2}+c_{3}\mathcal{L}_{R^{2}})\ ,\label{SUGRA}\\
	\mathcal{L}_{0} & =R+12\ell_{0}^{-2}-\frac{1}{4g_{0}^{2}}F^{\mu\nu}F_{\mu\nu}+\frac{1}{12\sqrt{3}g_{0}^{3}}\epsilon^{\mu\nu\rho\sigma\delta}F_{\mu\nu}F_{\rho\sigma}A_{\delta}\ ,\nonumber 
\end{align}
where $g_0$ is a bookkeeping parameter introduced by $A_\m\rightarrow A_\m/g_0$ and the 4-derivative corrections contain three parts $\mathcal{L}_{{\rm Weyl}^{2}}$,
$\mathcal{L}_{{\rm Ricci}^2}$ and $\mathcal{L}_{R^{2}}$. The Weyl squared
action $\mathcal{L}_{{\rm Weyl}^{2}}$ is
\begin{align}
	\mathcal{L}_{{\rm Weyl}^{2}} & =R_{\mu\nu\rho\sigma}R^{\mu\nu\rho\sigma}-\frac{4}{3}R_{\mu\nu}R^{\mu\nu}+\frac{1}{3}R^{2}-\frac{1}{2g_{0}^{2}}R_{\mu\nu\rho\sigma}F^{\mu\nu}F^{\rho\sigma}\nonumber \\
	& %%%
	-\frac{4}{3g_{0}^{2}}R_{\mu\nu}F^{\mu\lambda}F_{\ \lambda}^{\nu}+\frac{2}{9g_{0}^{2}}RF^{\mu\nu}F_{\mu\nu}-\frac{2}{9g_{0}^{2}}\nabla^{\mu}F_{\mu\rho}\nabla_{\nu}F^{\nu\rho}-\frac{61}{432g_{0}^{4}}(F^{\mu\nu}F_{\mu\nu})^{2}\nonumber \\
	& %%%
	+\frac{5}{8g_{0}^{4}}F_{\mu\nu}F^{\nu\lambda}F_{\lambda\delta}F^{\delta\mu}+\frac{5}{72\sqrt{3}g_{0}^{3}}\epsilon_{\mu\nu\rho\sigma\alpha}F^{\mu\nu}F^{\rho\sigma}\nabla_{\beta}F^{\beta\alpha}+\frac{200}{3\ell_{0}^{4}}-\frac{35}{9\ell_{0}^{2}g_{0}^{2}}F^{\mu\nu}F_{\mu\nu}\nonumber \\
	& %%%
	+\frac{20}{3\ell_{0}^{2}}R-\frac{1}{\sqrt{3}\ell_{0}^{2}g_{0}^{3}}\epsilon^{\mu\nu\rho\sigma\alpha}F_{\mu\nu}F_{\rho\sigma}A_{\alpha}+\frac{1}{2\sqrt{3}g_{0}}\epsilon^{\mu\nu\rho\sigma\alpha}A_{\mu}R_{\nu\rho}^{\ \ \ \beta\gamma}R_{\sigma\alpha\beta\gamma}\ ,
\end{align}
The Ricci tensor squared action $\mathcal{L}_{{\rm Ricci}^2}$ is
\begin{align}
	\mathcal{L}_{{\rm Ricci}^2} & =R^{2}-4R_{\mu\nu}R^{\mu\nu}-\frac{5}{6g_{0}^{2}}RF^{\mu\nu}F_{\mu\nu}-\frac{2}{g_{0}^{2}}\nabla^{\mu}F_{\mu\rho}\nabla_{\nu}F^{\nu\rho}+\frac{4}{g_{0}^{2}}R_{\mu\nu}F^{\mu\lambda}F_{\ \lambda}^{\nu}\nonumber \\
	& %%%
	+\frac{5}{16g_{0}^{4}}(F^{\mu\nu}F_{\mu\nu})^{2}-\frac{11}{9g_{0}^{4}}F_{\mu\nu}F^{\nu\lambda}F_{\lambda\delta}F^{\delta\mu}-\frac{2}{3\sqrt{3}g_{0}^{3}}\epsilon_{\mu\nu\rho\sigma\alpha}F^{\mu\nu}F^{\rho\sigma}\nabla_{\beta}F^{\beta\alpha}\nonumber \\
	& %%%
	-\frac{112}{\ell_{0}^{4}}+\frac{22}{3\ell_{0}^{2}g_{0}^{2}}F^{\mu\nu}F_{\mu\nu}-\frac{16}{\ell_{0}^{2}}R-\frac{\sqrt{3}}{\ell_{0}^{2}g_{0}^{3}}\epsilon^{\mu\nu\rho\sigma\alpha}A_{\mu}F_{\nu\rho}F_{\sigma\alpha}\ ,
\end{align}
The Ricci scalar squared action $\mathcal{L}_{R^{2}}$ is
\begin{align}
	\mathcal{L}_{R^{2}} & =R^{2}-\frac{1}{6g_{0}^{2}}RF^{\mu\nu}F_{\mu\nu}+\frac{1}{144g_{0}^{4}}(F^{\mu\nu}F_{\mu\nu})^{2}+\frac{16}{\ell_{0}^{2}}R+\frac{14}{3\ell_{0}^{2}g_{0}^{2}}F^{\mu\nu}F_{\mu\nu}-\frac{\sqrt{3}}{\ell_{0}^{2}g_{0}^{3}}\epsilon^{\mu\nu\rho\sigma\alpha}F_{\mu\nu}F_{\rho\sigma}A_{\alpha}+\frac{208}{\ell_{0}^{4}}\ .
\end{align}
To arrive at the action used in the body of the paper, we first perform the field redefinitions
\begin{align}
	g_{\mu\nu} & \rightarrow g_{\mu\nu}'=g_{\mu\nu}+\lambda_{0}g_{\mu\nu}+\lambda_{1}R_{\mu\nu}+\lambda_{2}g_{\mu\nu}R+\frac{\lambda_{3}}{g_{0}^{2}}F_{\mu\nu}^{2}+\frac{\lambda_{4}}{g_{0}^{2}}g_{\mu\nu}F^{2}\ ,\nonumber \\
	A_{\nu} & \rightarrow A_{\nu}'=A_{\nu}+\lambda_{5}A_{\nu}+\lambda_{6}\nabla^{\rho}F_{\rho\nu}+\frac{\lambda_{7}}{g_{0}}\epsilon_{\nu\alpha\beta\rho\lambda}F^{\alpha\beta}F^{\rho\lambda}\ ,
\end{align}
with the following parameters
\begin{align}
	& \lambda_{0}=4(\lambda_{1}+5\lambda_{5})\ell_{0}^{-2}=-\frac{4}{9}\big(5c_{1}+6(c_{2}+5c_{3})\big)\ell_{0}^{-2},\ \  \lambda_{1}=-4c_{2},\ \ \lambda_{2}=\frac{1}{9}\big(-c_{1}+6(c_{2}-c_{3})\big),\ \nonumber \\
	& \lambda_{3}=2(c_{2}-c_{1}),\ \ \lambda_{4}=\frac{1}{108}\left(49c_{1}+6(c_{3}-7c_{2})\right),\ \ \lambda_{5}=0,\ \ \lambda_{6}=\frac{2}{9}(c_{1}+9c_{2}),\ \ \lambda_{7}=\frac{4c_{2}-3c_{1}}{24\sqrt{3}}\ .
\end{align}
Consequently the effective action becomes 
\begin{align}
	I_{5D, N=1}= & \frac{\sigma}{16\pi}\int\sqrt{-g}d^{5}x(\mathcal{L}_{2\partial}+c_{1}\mathcal{L}_{\text{Weyl}^{2}})\ ,\ \mathcal{L}_{2\partial}=R+12\ell^{-2}-\frac{1}{4g^{2}}F^{\mu\nu}F_{\mu\nu}+\frac{\epsilon^{\mu\nu\rho\sigma\delta}F_{\mu\nu}F_{\rho\sigma}A_{\delta}}{12\sqrt{3}g^{3}}\ ,\label{EW}\\
	\mathcal{L}_{\text{Weyl}^{2}}= & -\frac{2}{g^{2}\ell^{2}}F^{\mu\nu}F_{\mu\nu}-\frac{\epsilon^{\mu\nu\rho\sigma\delta}F_{\mu\nu}F_{\rho\sigma}A_{\delta}}{\sqrt{3}g^{3}\ell^{2}}+C_{\mu\nu\rho\sigma}C^{\mu\nu\rho\sigma}-\frac{1}{2g^{2}}C_{\mu\nu\rho\sigma}F^{\mu\nu}F^{\rho\sigma}\nonumber \\
	& +\frac{13}{96g^{4}}(F^{\mu\nu}F_{\mu\nu})^{2}-\frac{13}{24g^{4}}F_{\mu\nu}F^{\nu\lambda}F_{\lambda\delta}F^{\delta\mu}+\frac{1}{2\sqrt{3}g}\epsilon^{\mu\nu\rho\sigma\alpha}A_{\mu}C_{\nu\rho}^{\ \ \ \beta\gamma}C_{\sigma\alpha\beta\gamma}\ ,\nonumber 
\end{align}
where
\begin{align}
	& 	{\sigma}=\sigma_0-24\s_0(c_{2}+c_{3})\ell_{0}^{-2},\quad {g}=g_0\big(1+4(c_{2}+c_{3})\ell_{0}^{-2}\big),
	\quad {\ell}=\ell_{0}\big(1-4(c_{2}+c_{3})\ell_{0}^{-2}\big)\ ,\label{EffParameter}
\end{align}
Here, it's important to emphasize that $\ell_{0}$, $g_{0}$ and $\sigma_{0}$
represent the bare physical parameters before the field redefinitions,
$\ell$, $g$ and $\sigma$ denote the effective parameters
following the field redefinitions, they encode contributions
from parameters of the original the 4-derivative supergravity
action (\ref{SUGRA}), such as $\ell_{0},\ g_{0},\ \sigma_{0}$ and
$c_{2},\ c_{3}$.

\subsection{C. Thermodynamic quantities}
The on-shell action of the general charged rotating AdS black holes \cite{{Chong:2005hr}} in $5D$ minimal gauged supergravity \eqref{5dsugra} is given by
\bea
I_{\rm ren}(T,\Phi_e,\Omega_a,\Omega_b)=\frac{\pi\beta\sigma}{4\Xi_{a}\Xi_{b}}\big(m-\frac{q^{2}r_{0}^{2}}{X+abq}-X\ell^{-2}\big)-\frac{c_{1}\pi\beta\sigma\Sigma_{k=0}^{8}d_{2k}r_{0}^{2k}}{4r_{0}^{4}\ell^{4}\Xi_{a}\Xi_{b}X\big(X+abq\big)}\ ,
\label{FullAction}
\eea
where $m=2r_{0}^{-2}\big(2abq+q^{2}+X(1+r_{0}^{2}\ell^{-2})\big),\ X=(a^{2}+r_{0}^{2})(b^{2}+r_{0}^{2}),\ \Xi_{a}=1-a^2\ell^{-2},\ \Xi_{b}=1-b^2\ell^{-2},\ \beta=1/T,\  y=\frac{q}{ab},\ Y_{m,n}=(ab)^{m}(a^{n}+b^{n})$ and
\bea
d_{0}&=&\frac{1}{2}Y_{6,0}\ell^{4}(y+1)^{5},\ \ \ d_{2}=\ell^{2}(y+1)^{3}(Y_{6,0}-\ell^{2}(3y+5)Y_{4,2})\ , \cr
%%%
d_{4}&=&\frac{1}{2}(y+1)\big[Y_{6,0}-4(3y^{2}+9y+5)\ell^{2}Y_{4,2}-\ell^{4}\big((3y^{3}+19y^{2}+45y+31)Y_{4,0}+2(2y+5)Y_{2,4}\big)\big]\ ,\cr
%%%
d_{6}&=&\ell^{4}\big(Y_{0,6}-(8y^{2}+26y+23)Y_{2,2}\big)-2\ell^{2}\big[(5-y^{2}+y)Y_{2,4}+\frac{1}{2}\big(31-y(y-26)(y+2)\big)Y_{4,0}\big]-5(2y+1)Y_{4,2}\ ,\cr
%%%
d_{8}&=&\ell^{4}\big(3Y_{0,4}-\frac{1}{2}(6y^{2}+3y+7)Y_{2,0}\big)+2\ell^{2}\big((4y^{2}-6y-23)Y_{2,2}+Y_{0,6}\big)+5(y-1)Y_{2,4}-\frac{1}{2}(28y+31)Y_{4,0}\ , \cr
%%%
d_{10}&=&11\ell^{4}Y_{0,2}+\ell^{2}\big((3y^{2}+11y-7)Y_{2,0}+6Y_{0,4}\big)+(14y-23)Y_{2,2}+Y_{0,6}\ , \cr
%%%
d_{12}&=&9\ell^{4}+22\ell^{2}Y_{0,2}+\frac{1}{2}(25y-7)Y_{2,0}+3Y_{0,4}\ ,\ \ \ d_{14}=18\ell^{2}+11Y_{0,2}\ ,\ \ \ d_{16}=9\ ,
\eea
where the effects of supersymmetric Ricci tensor squared and Ricci scalar squared are encoded in the effective AdS radius $\ell$ indicated in \eqref{tp}.
The other thermodynamic
variables are obtained from the action \eqref{FullAction} via standard relations
\bea
I_{\text{ren}}&=&\beta G=\beta(M-TS-\Phi_{e}Q_{e}-\Omega_{a}J_{a}-\Omega_{b}J_{b})\ ,\cr
%%%
S&=&-\frac{\partial G}{\partial T}|_{\Phi_{e},\Omega_{a,b}},\ Q_{e}=-\frac{\partial G}{\partial\Phi_{e}}|_{T,\Omega_{a,b}},\ J_{a(b)}=-\frac{\partial G}{\partial\Omega_{a(b)}}|_{T,\Phi_{e},\Omega_{b(a)}}\ ,
\eea
where $\beta, \ \Phi_e, \ \Omega_{a,b}$ are listed in (\ref{Tem},\ref{tp}) and  $G$ denotes the Gibbs free energy.

Next,  we consider the supersymmetric and extremal (BPS) limit of the entropy and  conserved charges. The limiting procedure for the corrected BPS charges is  not unique \cite{Cabo-Bizet:2018ehj}. We will present the BPS charges obtained using two different limiting procedures. The first approach involves applying
\bea
q\rightarrow-ab+r_{0}^{2}\ell^{-1}\sqrt{a^{2}+b^{2}+\ell^{2}+2r_{0}^{2}}\  ,
\eea
followed by $r_{0}\rightarrow\sqrt{ab+\ell(a+b)}$, resulting in the following corrected BPS charges
%Taking the BPS limit, the conserved charges become
\bea
M_{*}&=&-\frac{\pi\sigma\ell^{2}(a+b)\left(\ell(a^{2}-ab+b^{2})+ab(a+b)-3\ell^{3}\right)}{4(a-\ell)^{2}(b-\ell)^{2}}-\frac{2\pi c_{1}\sigma}{(a+b)(a-\ell)^{2}(a+\ell)(b-\ell)^{2}(b+\ell)}\Big[ \cr
%%%
&-&\ell^{4}(a+b)(7a^{2}+10ab+7b^{2})+2\ell^{3}\big((a^{2}+b^{2})(a^{2}-10ab+b^{2})-20a^{2}b^{2}\big)+6\ell^{2}(a-b)^{2}(a+b)^{3}\cr
%%%
&+&\ell\big(4a^{3}b^{3}+(a+b)^{2}(a^{2}+b^{2})(a^{2}+8ab+b^{2})\big)+(a^{2}+ab+b^{2})\left(ab(a+b)(a^{2}+6ab+b^{2})-4\ell^{5}\right)\Big]\ ,\cr
%%%
S_{*}&=&\frac{\pi^{2}\sigma\ell^{3}(a+b)\sqrt{\ell(a+b)+ab}}{2(a-\ell)(b-\ell)}+\frac{2\pi^{2}c_{1}\sigma\ell}{(a+b)(a^{2}-\ell^{2})(b^{2}-\ell^{2})\sqrt{\ell(a+b)+ab}}\Big[\cr
%%%
&+&\ell^{4}(5a^{2}+2ab+5b^{2})+12\ell^{3}(a+b)(a^{2}+b^{2})+ab(a+b)^{2}(a^{2}+7ab+b^{2})\cr
%%%
&+&\ell^{2}\big(8(a+b)^{2}(a^{2}+\frac{5ab}{4}+b^{2})-8a^{2}b^{2}\big)+\ell(a+b)^{3}(a^{2}+10ab+b^{2})\Big]\ ,\cr
%%%
Q_{e,*}&=&\frac{\sqrt{3}\pi\sigma\ell^{3}(a+b)}{4g(a-\ell)(b-\ell)}+\frac{2\pi c_{1}\sigma}{\sqrt{3}g(a+b)(a^{2}-\ell^{2})(b^{2}-\ell^{2})}\Big[\ell^{2}(a+b)(7a^{2}+10ab+7b^{2}) \cr
%%%
&+&4\ell^{3}(a^{2}+ab+b^{2})+\ell\big((a+b)^{2}(a^{2}+14ab+b^{2})-4a^{2}b^{2}\big)+ab(a+b)(a^{2}+10ab+b^{2})\Big]\ ,\cr
%%%
J_{a,*}&=&-\frac{\pi\sigma\ell^{3}(a+b)\big(\ell(2a+b)+ab\big)}{4(a-\ell)^{2}(b-\ell)}-\frac{2\pi c_{1}\sigma\ell}{(a+b)(a+\ell)(a-\ell)^{2}(b^{2}-\ell^{2})}\Big[-2\ell^{5}(a-b)\cr
%%%
&-&2\ell^{4}(a-b)(3a+4b)+2\ell^{3}(a^{3}-3a^{2}b+3ab^{2}+5b^{3})+\ell^{2}(a^{2}+b^{2})(7a^{2}+16ab+7b^{2})\cr
%%%
&+&\ell(a+b)(a^{4}+14a^{3}b+16a^{2}b^{2}+10ab^{3}+b^{4})+ab(a+b)(a^{3}+9a^{2}b+7ab^{2}+b^{3})\Big]\ .
\label{Method1}
\eea
and $J_{b,*}=J_{a,*}|_{a\leftrightarrow b}$. We can verify that these BPS charges  satisfy non-linear relations
\bea
&& \big(3Q_{R}+4(2\mathrm{a}-\mathrm{c})\big)\big(3Q_{R}^{2}-8\mathrm{c}(J_{a,*}+J_{b,*})\big)\nonumber \\
&=& Q_{R}^{3}+16(3\mathrm{c}-2\mathrm{a})J_{a,*}J_{b,*}+64\mathrm{a}(\mathrm{a}-\mathrm{c})\frac{(Q_{R}+\mathrm{a})(J_{a,*}-J_{b,*})^{2}}{Q_{R}^{2}-2\mathrm{a}(J_{a,*}+J_{b,*})}\ .
\label{non-linear}
\eea
The second approach is used in \cite{Cassani:2022lrk,Cabo-Bizet:2018ehj}
\be
q\rightarrow\ell^{-1}(a+b)(a+\ell)(b+\ell)\ ,\ \ \ r_{0}\rightarrow\sqrt{ab+\ell(a+b)}+\epsilon\ ,
\ee
where $\epsilon$ is the expansion parameter and it yields the following  BPS charges
\bea
M_{*}&=&-\frac{\pi\sigma\ell^{2}(a+b)\left(\ell(a^{2}-ab+b^{2})+ab(a+b)-3\ell^{3}\right)}{4(a-\ell)^{2}(b-\ell)^{2}}-\frac{2\pi c_{1}\sigma}{(a+b)(a-\ell)^{2}(a+\ell)(b-\ell)^{2}(b+\ell)Z}\Big[2\ell^{6}(a+b)\big( \cr
%%%	
&+&51a^{4}+329a^{3}b+593a^{2}b^{2}+329ab^{3}+51b^{4}\big)+\ell^{2}\big(a^{9}-68a^{7}b^{2}-464a^{6}b^{3}-1122a^{5}b^{4}-1122a^{4}b^{5}+b^{9} \cr
%%%	
&-&464a^{3}b^{6}-68a^{2}b^{7}\big)+ab\ell(2a^{8}+a^{7}b-40a^{6}b^{2}-208a^{5}b^{3}-350a^{4}b^{4}-208a^{3}b^{5}-40a^{2}b^{6}+ab^{7}+2b^{8})\cr
%%%	
&+&\ell^{9}(23a^{2}+50ab+23b^{2})-\ell^{4}(a+b)(7a^{6}+102a^{5}b+335a^{4}b^{2}+498a^{3}b^{3}+335a^{2}b^{4}+102ab^{5}+7b^{6})\cr
%%%	
&+&9\ell^{8}(a+b)(9a^{2}+23ab+9b^{2})+2\ell^{5}(5a^{6}+106a^{5}b+423a^{4}b^{2}+648a^{3}b^{3}+423a^{2}b^{4}+106ab^{5}+5b^{6})\cr
%%%	
&+&2\ell^{7}(70a^{4}+361a^{3}b+590a^{2}b^{2}+361ab^{3}+70b^{4})+a^{2}b^{2}(a^{7}-9a^{5}b^{2}-31a^{4}b^{3}-31a^{3}b^{4}-9a^{2}b^{5}+b^{7})\cr
%%%	
&-&\ell^{3}(a^{8}+44a^{7}b+414a^{6}b^{2}+1350a^{5}b^{3}+1974a^{4}b^{4}+1350a^{3}b^{5}+414a^{2}b^{6}+44ab^{7}+b^{8})+3\ell^{10}(a+b)\Big]\ , \cr
%%%	
S_{*}&=&\frac{\pi^{2}\sigma\ell^{3}(a+b)\sqrt{\ell(a+b)+ab}}{2(a-\ell)(b-\ell)}-\frac{4\pi^{2}c_{1}\sigma\ell\sqrt{a(b+\ell)+b\ell}}{(a+b)(a^{2}-\ell^{2})(b^{2}-\ell^{2})Z}\Big[3\ell^{5}(a+b)(15a^{2}+38ab+15b^{2})\cr
%%%	
&+&\ell^{4}(59a^{4}+313a^{3}b+504a^{2}b^{2}+313ab^{3}+59b^{4})-ab(a^{6}-12a^{4}b^{2}-26a^{3}b^{3}-12a^{2}b^{4}+b^{6}) \cr
%%%	
&+&\ell^{3}(a+b)^{3}(23a^{2}+176ab+23b^{2})+\ell^{2}(a^{6}+63a^{5}b+312a^{4}b^{2}+496a^{3}b^{3}+312a^{2}b^{4}+63ab^{5}+b^{6})\cr
%%%	
&+&\ell^{6}(11a^{2}+26ab+11b^{2})-\ell(a+b)(a^{6}-2a^{5}b-46a^{4}b^{2}-110a^{3}b^{3}-46a^{2}b^{4}-2ab^{5}+b^{6})\Big]\ ,\cr
%%%	
Q_{e,*}&=&	\frac{\sqrt{3}\pi\sigma\ell^{3}(a+b)}{4g(a-\ell)(b-\ell)}+\frac{2\pi c_{1}\sigma}{\sqrt{3}g(a+b)(a^{2}-\ell^{2})(b^{2}-\ell^{2})Z}\Big[-3\ell^{8}(a+b)-2\ell^{6}(a+b)(31a^{2}+79ab+31b^{2})\cr
%%%	
&-&4\ell^{5}(25a^{4}+129a^{3}b+211a^{2}b^{2}+129ab^{3}+25b^{4})-2\ell^{4}(a+b)(a^{2}+4ab+b^{2})(40a^{2}+93ab+40b^{2})\cr
%%%	
&-&4\ell^{7}(5a^{2}+11ab+5b^{2})-2\ell^{3}\big(13a^{6}+161a^{5}b+571a^{4}b^{2}+844a^{3}b^{3}+571a^{2}b^{4}+161ab^{5}+13b^{6}\big)\cr
%%%	
&-&\ell^{2}(a+b)\big(6a^{2}b^{2}(26a^{2}+49ab+26b^{2})+(a^{2}+52ab+b^{2})(a+b)^{4}\big)+2ab\ell\big(2(a^{2}-17ab+b^{2})(a+b)^{4}\cr
%%%	
&+&3ab(3a^{4}-4a^{2}b^{2}+3b^{4})\big)+a^{2}b^{2}(a+b)(5a^{4}-12a^{3}b-43a^{2}b^{2}-12ab^{3}+5b^{4})\Big]\ ,\cr
%%%	
J_{a,*}	&=&-\frac{\pi\sigma\ell^{3}(a+b)\big(\ell(2a+b)+ab\big)}{4(a-\ell)^{2}(b-\ell)}+\frac{2\pi c_{1}\sigma\ell}{(a+b)(a-\ell)^{2}(a+\ell)(b^{2}-\ell^{2})Z}\Big[4\ell^{7}(2a+b)(a+2b)(3a+2b)\cr
%%%	
&+&2\ell^{6}(a+2b)(33a^{3}+81a^{2}b+59ab^{2}+13b^{3})+2\ell^{5}(a+2b)(3a+2b)(21a^{3}+52a^{2}b+39ab^{2}+6b^{3})\cr
%%%	
&+&\ell^{8}(a+b)(5a+b)+\ell^{2}(a^{8}+68a^{7}b+476a^{6}b^{2}+1086a^{5}b^{3}+1018a^{4}b^{4}+402a^{3}b^{5}+58a^{2}b^{6}-b^{8})\cr
%%%	
&+&2\ell^{4}(50a^{6}+377a^{5}b+917a^{4}b^{2}+1013a^{3}b^{3}+537a^{2}b^{4}+118ab^{5}+6b^{6})+2a\ell^{3}\big(196a^{5}b+707a^{4}b^{2}\cr
%%%	
&+&15a^{6}+1032a^{3}b^{3}+699a^{2}b^{4}+214ab^{5}+23b^{6}\big)-2ab\ell(a^{7}-19a^{6}b-111a^{5}b^{2}-184a^{4}b^{3}-109a^{3}b^{4}\cr
%%%	
&-&18a^{2}b^{5}+ab^{6}+b^{7})+a^{2}b^{2}(-3a^{6}+4a^{5}b+38a^{4}b^{2}+48a^{3}b^{3}+12a^{2}b^{4}-2ab^{5}-b^{6})\Big]\ ,
\label{Method2}
\eea
where $J_{b,*}=J_{a,*}|_{a\leftrightarrow b}$ and
\bea
Z=a^{4}-2a^{3}b-9a^{2}b^{2}-2ab^{3}+b^{4}-2(a+b)\left(a^{2}+11ab+b^{2}\right)\ell-3\ell^{2}(3a^{2}+8ab+3b^{2})-2\ell^{3}(a+b)+\ell^{4} \ .
\eea
Again, these BPS charges satisfy the  non-linear relations \eqref{non-linear}. Interestingly, if we apply the following redefinition of parameters $a,b$  to  \eqref{Method2}
\bea
a\rightarrow a'=a+\delta a\ ,\ \ b\rightarrow b'=b+\delta b\ ,
\eea
\bea
\delta a&=&-\frac{8c_{1}(a-\ell)}{3Z\ell^{3}(a+b)^{2}(a+\ell)(b+\ell)^{2}}\Big[a^{8}b^{2}+4a^{7}b^{3}-13a^{6}b^{4}-67a^{5}b^{5}-67a^{4}b^{6}-13a^{3}b^{7}+4a^{2}b^{8}+ab^{9} \cr
%%%
&-&6\ell^{9}(a-b)+3b\ell^{8}(a+b)+\ell^{4}(54a^{6}+595a^{5}b+1274a^{4}b^{2}+513a^{3}b^{3}-633a^{2}b^{4}-460ab^{5}-47b^{6})\cr
%%%
&+&12\ell^{7}(8a^{3}+13a^{2}b-7ab^{2}-6b^{3})+\ell^{5}(162a^{5}+975a^{4}b+1162a^{3}b^{2}-176a^{2}b^{3}-572ab^{4}-111b^{5})\cr
%%%
&+&\ell^{6}(198a^{4}+691a^{3}b+221a^{2}b^{2}-409ab^{3}-137b^{4})+\ell^{3}\big(6a^{7}+157a^{6}b+516a^{5}b^{2}+279a^{4}b^{3}-602a^{3}b^{4}\cr
%%%
&-&729a^{2}b^{5}-204ab^{6}+b^{7}\Big)+b\ell^{2}(18a^{7}+79a^{6}b-62a^{5}b^{2}-623a^{4}b^{3}-797a^{3}b^{4}-320a^{2}b^{5}-17ab^{6}+6b^{7})\cr
%%%
&+&b\ell(a^{8}+8a^{7}b-21a^{6}b^{2}-236a^{5}b^{3}-440a^{4}b^{4}-254a^{3}b^{5}-27a^{2}b^{6}+8ab^{7}+b^{8})\Big]\ ,\cr
%%%
\delta b	&=&\delta a|_{a\leftrightarrow b}\ .
\eea
we recover the results given in \eqref{Method1}. In the case where $a=b$  in \eqref{Method2}, similar results were obtained in \cite{Cassani:2022lrk} where it is the bare AdS radius rather than the effective AdS radius that enters the BPS charges.

\end{document}